\documentclass[%
 reprint,
superscriptaddress,
 amsmath,amssymb,
 aps,
]{revtex4-2}

	\usepackage{bm}
	\usepackage{braket}
	\usepackage{dsfont}
 	\usepackage{graphicx}
	\usepackage[caption=false]{subfig}
	\usepackage[export]{adjustbox}
	\usepackage{tabularx}

	\usepackage{xcolor,colortbl}
\usepackage[english]{babel}
\usepackage{amsmath}
\usepackage{amssymb, comment}
\usepackage[]{graphicx}
\usepackage{times}
\usepackage{bm}
\usepackage{units}
\usepackage{braket}
\usepackage[version=3]{mhchem}

\usepackage{bbm}
\usepackage{tabularx}

\usepackage{color}
	\newcommand{\vect}[1]{\boldsymbol{#1}}		% I use command \vect{} for vectors
	\newcommand{\op}[1]{\hat{\boldsymbol{#1}}}	% I use command \op{} for operators
	

\begin{document}

\title{Singlet exciton optics and phonon-mediated dynamics in oligoacene semiconductor crystals}

\author{Joshua J. P. Thompson}
\email{joshua.thompson@physik.uni-marburg.de
}
\affiliation{Department of Physics, Philipps-Universit\"{a}t Marburg, Renthof 7, 35032 Marburg}

\author{Dominik Muth}
\affiliation{Department of Physics, Philipps-Universit\"{a}t Marburg, Renthof 7, 35032 Marburg}

\author{Sebastian Anh\"{a}user}
\affiliation{Department of Physics, Philipps-Universit\"{a}t Marburg, Renthof 7, 35032 Marburg}
\author{Daniel Bischof}
\affiliation{Department of Physics, Philipps-Universit\"{a}t Marburg, Renthof 7, 35032 Marburg}
\author{Marina Gerhard}
\affiliation{Department of Physics, Philipps-Universit\"{a}t Marburg, Renthof 7, 35032 Marburg}
\author{Gregor Witte}
\affiliation{Department of Physics, Philipps-Universit\"{a}t Marburg, Renthof 7, 35032 Marburg}
\author{Ermin Malic}
\affiliation{Department of Physics, Philipps-Universit\"{a}t Marburg, Renthof 7, 35032 Marburg}

\date{\today}

\begin{abstract}
Organic semiconductor crystals stand out as an efficient, cheap and diverse platform for realising optoelectronic applications. The optical response of these crystals is governed by a rich tapestry of exciton physics. So far, little is known on the phonon-driven singlet exciton dynamics in this class of materials.
In this joint theory-experiment work, we combine the fabrication of a high-quality oligoacene semiconductor crystal and characterization via photoluminescence measurements with a sophisticated approach to the microscopic modeling in these crystals. This allows us to investigate singlet exciton optics and dynamics. We  predict phonon-bottleneck effects in pentacene crystals, where we find dark excitons acting as crucial phonon-mediated relaxation scattering channels. While the efficient singlet fission in pentacene crystals hampers the experimental observation of this bottleneck effect,  we reveal both in theory and experiment a distinct polarisation- and temperature-dependence in absorption and photoluminescence spectra of tetracene crystals, including microscopic origin of exciton linewidths, the activation of the higher Davydov states at large temperatures, and polarisation-dependent quenching of specific exciton resonances.  Our joint theory-experiment study represents a significant advance in microscopic understanding of singlet exciton optics and dynamics in oligoacene crystals.
\end{abstract}

\maketitle
%\section{Introduction}

Organic semiconductor crystals  (OSCs) represent promising candidates in the development and design of new photovoltaics \cite{wilson2011ultrafast, bettis2017ultrafast, congreve2013external}, field effect transistors \cite{muccini2006bright, schweicher2020molecular} and biological sensors \cite{brega2020acenes}. This technological promise stems from a multitude of factors including  high optical absorption, efficient modification through functionalisation \cite{bischof2022regioselective} and  changes in the growth conditions \cite{siegrist2001enhanced, mattheus2003modeling} as well as cheap fabrication \cite{huang2018organic}. 
Of particular interest are optical properties of larger oligoacene crystals, such as tetracene \cite{grumstrup2010enhanced, tayebjee2013exciton} and pentacene \cite{he2005fundamental, dressel2008kramers} crystals, which serve as prototypical materials. Their properties are dictated by the formation of tightly bound excitons (Coulomb-bound electron-hole pairs). In early studies on oligoacene crystals, it has been proposed that the optical properties are governed by strongly localized Frenkel excitons \cite{kasha1965exciton, yamagata2011nature}, where the bound electrons and holes are located on the same molecule. In pentacene and tetracene OSCs, however, the Frenkel exciton model breaks down \cite{schuster2007exciton, gunder2021polarization}, where, notwithstanding the weak intermolecular coupling \cite{de2003anisotropy, cheng2003three}, excitons are distributed across several molecules \cite{schuster2007exciton,cocchi2018polarized, sharifzadeh2013low}.  These excitons represent a middle ground between  extremely localised Frenkel excitons with large binding energies and weaker bound, spatially delocalised Wannier excitons observed in covalently-bound systems, such as transition metal dichalcogenides \cite{chernikov2014exciton, mueller2018exciton}. 

Tetracene and pentacene  are particularly important due to the singlet-fission \cite{wilson2011ultrafast, wilson2013singlet, wu2014singlet, huang2018organic, grumstrup2010enhanced}, whereby a singlet exciton decays into two triplet excitons. This charge-transfer driven process is only likely if the energy of two triplet excitons is lower or similar than the energy of a singlet exciton. In tetracene and pentacene, this condition is fulfilled and singlet excitons readily decay into two triplet states \cite{zimmerman2010singlet, wu2014singlet}, occurring on an ultrafast timescale (fs) in pentacene \cite{wilson2011ultrafast}, and more slowly in tetracene (ps) \cite{wilson2013temperature}.  For photovoltaics applications this is extremely useful as, unlike the singlet excitons formed when sunlight impinges on the crystal, the resultant triplet excitons have a long-lifetime \cite{tayebjee2013exciton} with the possibility of harvesting four charge carriers for a single incident photon and thus overcoming the Shockley-Queisser limit. This is important for minimising the loss and boosting the efficiency of photovoltaic devices.  

The dynamics of the excitonic system is dependent on a multitude of interaction mechanisms, such as exciton-phonon scattering. Hence, it is imperative that the singlet-exciton dynamics in these systems is well understood. While there is a wealth of literature on exciton optics \cite{dimitriev2022dynamics} and the role of singlet fission \cite{smith2010singlet, smith2013recent}, there are few studies on phonon-mediated exciton dynamics \cite{seiler2021nuclear, morrison2017evidence}. 
Typically, these OSC systems are treated theoretically using either DFT \cite{coropceanu2002hole, hummer2005electronic, xie2018nonlocal, brown2020band},  GW \cite{tiago2003ab}, Bethe-Salpeter \cite{cocchi2018polarized, ambrosch2009role} or Holstein-like Hamiltonians \cite{ortmann2010charge, hestand2015polarized, duan2019ultrafast, morrison2017evidence}. While successful in recovering the purely excitation-induced phenomena, such as the Davydov splitting \cite{cocchi2018polarized, ambrosch2009role} and the vibronic progression \cite{hestand2015polarized}, which is often described by means of model Hamiltonians \cite{hestand2018expanded}, the role of phonons, in particular intermolecular phonons \cite{draxl2014organic, yang2017intermolecular} and the phonon-driven exciton dynamics is difficult to describe with these models.  To this end, it is necessary to use an alternative approach which can capture these processes in OSC systems. In this work, we use the Wannier equation and density matrix formalism to microscopically derive  optical and dynamical properties of pentacene and tetracene crystals.  We demonstrate in a joint theory-experiment study the role of phonons in the optical response of OSC and in particular how optically excited excitons relax along the molecular exciton landscape, elucidating the importance of phonon-scattering channels, and the conditions for a phonon-bottleneck effect.

We recover the Davydov-splitting, observed in experiments, and calculate polarisation-dependent absorption and photoluminescence (PL) spectra of pentacene OSC. We calculate the exciton dynamics and find that dark excitons act as crucial scattering channels between the two Davydov-split excitonic states.  We find that the flatness of excitonic states leads to a phonon-bottleneck effect and  this has  ramifications on the relative intensity of the PL emission between the Davydov excitons. Practically, singlet fission dynamics hampers the direct observation of the bottleneck effect in pentacene in PL measurements. Therefore, we fabricate high-quality tetracene crystals and perform polarization- and temperature-dependent absorption and PL measurements. We find very good agreement between experiment and theory in terms of temperature-dependent linewidths of excitonic resonances and their intensity ratio as well as polarization-dependent suppression/emergence of exciton resonances.  Our joint theory-experiment study sheds light on fundamental relaxation processes in the technologically promising oligoacene crystals, in particular emphasizing the importance of exciton-phonon scattering and demonstrating the strength of our theoretical approach.

\begin{figure}[!t]
    \centering
    \includegraphics[width=0.99\linewidth]{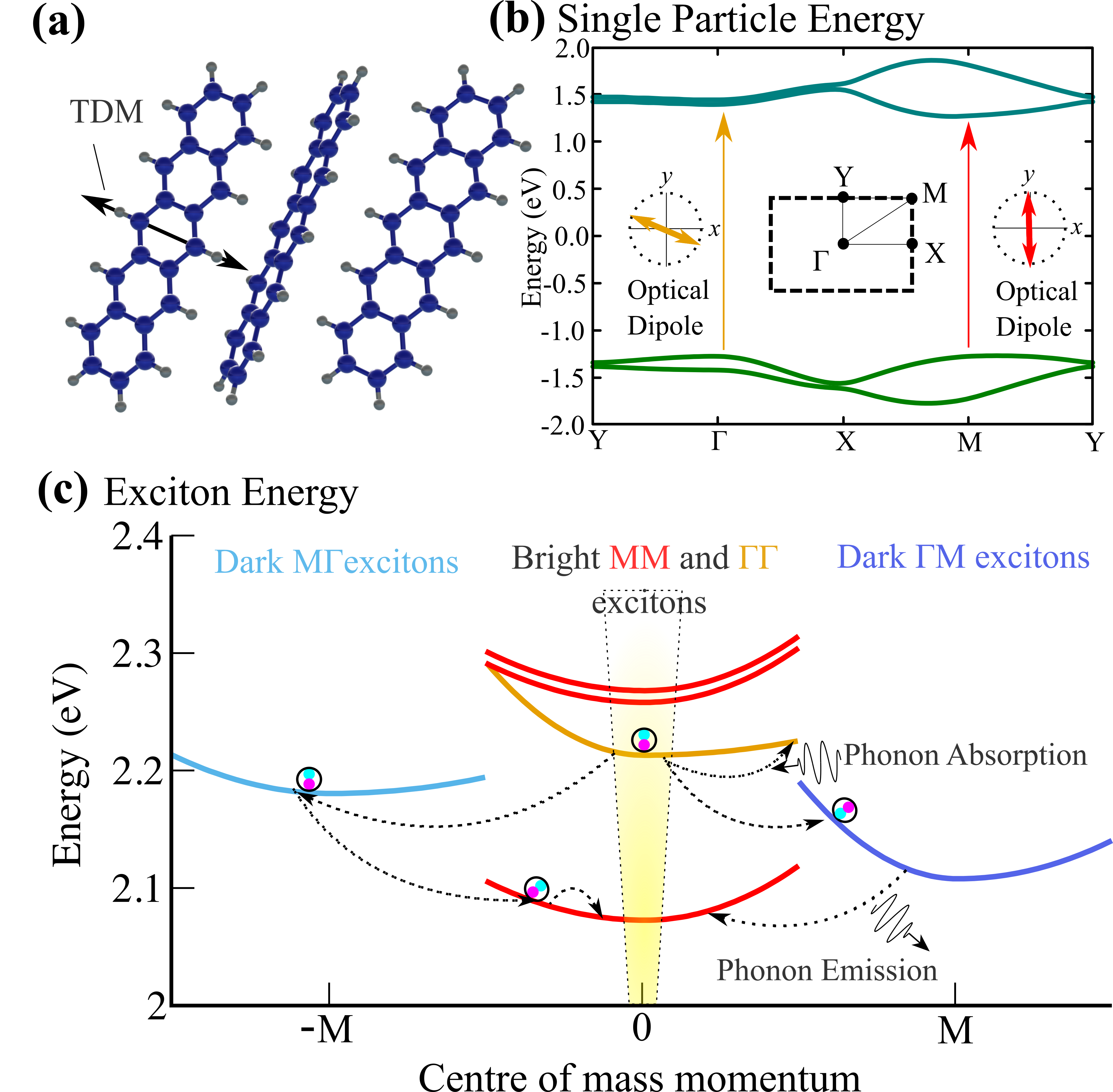}
    \caption{(a) Typical herringbone lattice structure of oligoacene crystals with the lowest monomer transition dipole moment shown in black. (b) Electronic dispersion of pentacene crystal for the HOMO and LUMO bands with two direct transitions at the M point (red) and at the $\Gamma$ point (orange). Insets show the relative dipole orientation of the two transitions as well as an in-molecular-plane slice of the  Brillouin zone. (c) The resulting excitonic dispersion including the momentum-dark  M$\Gamma$ and $\Gamma$M excitons. Following creation of an exciton in the light cone (yellow), exemplary scattering channels are indicated with dashed arrows and accompanied by emission/absorption of a phonon. }
    \label{fig:my_label1}
\end{figure}
\section{Results and Discussion}
\subsection{Fabrication, characterization, modeling}
To study exciton optics and dynamics in oligoacene crystals, we combine methods of a fully microscopic many-particle theory with fabrication of high-quality molecular crystals and polarization- and temperature-dependent absorption and PL measurements.

\textbf{Fabrication:} Fabrication
Ultrathin (001)-oriented tetracene crystals (d < 30 nm) were prepared by drop casting of toluene solutions onto quartz glass slides. To identify the azimuthal directions, larger millimeter-sized crystals were grown from solution that enable in-plane X-ray diffraction measurements and thereby allow correlating the orientation of the $\vect{a}$ and $\vect{b}$ unit cell vectors with the habit of the crystals as well as the polarization contrast of the photoluminescence (for details see Supplementary Information). The latter information then enabled a corresponding azimuth determination also for the ultrathin samples that were used for the optical analysis, and the homogeneous contrast in the PL micrographs confirmed the single crystals nature of such samples.

\textbf{Characterization:}
For the absorption and PL measurments, the samples were mounted in an evacuated helium flow cryostat. We probed the two Davydov MM and $\Gamma \Gamma$ excitons separately by positioning an analyzer in the detection path.  We recorded absorption and time resolved photoluminescence (TRPL) data for both orientations of the polariser (x and y). 
The absorption of the samples was probed using a broadband white light tungsten halogen lamp.  The thin crystals were sandwiched between two glass cover slips, to enable the transmission measurements. As a result, signatures of an etalon effect were observed in the raw data. Therefore, we pre-processed the absorption data using a FFT bandstop filter to remove the oscillation.
TRPL spectroscopy was conducted with a synchroscan streak camera. The excitation pulse duration was 100 fs with a wavelength of 440 nm. We distinguished the photoluminescence signal from the laser using a long pass filter.

\textbf{Modeling:}
The starting point for our calculation is  the electronic band structure of the OSC. In a crystal, tetracene and pentacene typically form a herringbone lattice, with their unit cell composed of two relatively rotated monomers (Fig \ref{fig:my_label1} (a)). While there are several different pentacene polymorphs \cite{ cocchi2018polarized}, in this work we present data for the Campbell polymorph \cite{de2003anisotropy}. The other well-known polymorphs,  such as the Siegrist and thin-film phase, have qualitatively and quantitatively similar optical properties \cite{ambrosch2009role, cocchi2018polarized}, such that our calculations can be generalised to these other structures.  The long axis of the monomer is mostly orthogonal to the layer plane, however both molecules are tilted.  
We use a tight-binding (TB) formalism \cite{de2003anisotropy, hummer2005electronic, guerrini2021long, cheng2003three} constructed from the highest-occupied-molecular-orbital (HOMO) and the lowest-unoccupied-molecular-orbital (LUMO) on the two molecules in the unit cell. Our resulting 4-orbital basis leads to an electronic  bandstructure where the intermolecular coupling lifts  the degeneracy of the HOMO  and LUMO bands of the herringbone OSC (cf. Fig \ref{fig:my_label1} (b)).  We extract TB parameters from literature \cite{de2003anisotropy} for pentacene, with good quantitative agreement with previous DFT and GW calculations \cite{hummer2005electronic, rangel2016structural,cudazzo2015exciton, cocchi2018polarized, mattheus2003modeling}.  

Importantly, we find two distinct direct band gaps at the M and $\Gamma$ valleys, marked with the red and orange arrows, respectively, in Fig \ref{fig:my_label1}.  
The transition dipole moment  of the lowest energy excitation of oligoacene monomers is typically oriented along the short-axis of the molecule (as indicated by the black arrow in  Fig. \ref{fig:my_label1} (a)).  In an OSC, the dipole moment of a given excitonic transition is determined by the linear combination of the molecular dipole moments originating from both symmetrically distinct molecules within the unit cell, according to the Kasha model \cite{kasha1965exciton}, with weighting determined by the TB wavefunctions \cite{pedersen2001optical, guerrini2021long}. 
Using our  TB model, we find the transition dipole moment at the M point to be oriented along the $y$-direction, while at the $\Gamma$ point it is oriented approximately $70^\circ$ from the $y$-axis, cf. the inset of Fig \ref{fig:my_label1} (b). Here we are describing only the dipole orientation within the layer, however due to the tilting of the molecules, the dipole also has a non-zero component in the $z$-direction at both the $\Gamma$ and the M point. 

Around these two distinct exciton valleys, we can use the Wannier equation to calculate the exciton binding energies and wavefunctions (see Supplementary Information for more details) \cite{kira2006many}.
 The resulting binding energies  $E^b_{ij, \mu}$ and excitonic wavefunctions resemble those of a 2D hydrogen atom with $\mu  \in (1s, 2s, 2p_x, ... )$ \cite{chernikov2014exciton}, albeit anisotropic, owing to the anisotropy of the underlying crystal structure \cite{usman2022enhanced}.
In addition to the exciton states formed around the direct band gap in the crystal, there are also momentum-dark excitons formed when the electron and hole originate from different valleys \cite{selig2016excitonic, berghauser2018mapping}, as shown in the excitonic dispersion in Fig 1(c).  The separation between the minima of the MM and $\Gamma\Gamma$ excitons is known as the Davydov splitting, and originates from the coupling between the two symmetrically distinct molecules in the unit cell. These two Davydov-split excitons carry specific transition dipoles and our estimate of 140 meV is in good agreement with previous calculations and experimental measurements on pentacene crystals \cite{cocchi2018polarized}.  

The behaviour and nature of excitons formed in OSCs can be probed via  absorption and PL measurements. To that end, we derive an expression for the optical absorption of OSCs. We use Heisenberg’s equation of motion \cite{kira2006many} to derive
the semiconductor luminescence equations \cite{brem2018exciton}, allowing us to obtain the Elliot formula for the optical absorption
\begin{align}
    I^\sigma_\text{abs}(\omega)  &=  \dfrac{2}{\hbar} \sum_{\mu, \xi} \text{Im}\dfrac{|{\tilde{M}}^{\sigma }_{\mu\xi}|^2}{E^b_{\xi\mu} - \hbar\omega - i(\gamma_{\xi \sigma}^\mu + \Gamma^{\mu}_\xi)}
\end{align}
where $\sigma$ is the optical polarisation of the absorbed light and  $\xi = (i,j)$ the  valley index  for the electron in valley $i$ and the hole in valley $j$. We define ${\tilde{M}}^{\sigma }_{\mu\xi}$ as the optical matrix element with  ${\tilde{M}}^{\sigma }_{\mu\xi} = \vect{d}_\xi \cdot \op{e}_\sigma \sum_{\vect{k}} \varphi^{\xi, \mu}_{\vect{k}}$. Here $\vect{d}_\xi$ is the optical dipole between the HOMO and LUMO bands.     
Furthermore,  $\gamma_{\xi \sigma}^\mu$ describes the broadening due to singlet exciton losses in our system. This is composed of the radiative decay and the singlet-triplet fission mechanisms. In these materials, the radiative decay rate is very low with lifetimes of  $\sim12$ ns  \cite{burdett2010excited, burdett2012quantum, camposeo2010polarized}. In contrast, the singlet-fission decay rate is much faster, with lifetimes as low as $\sim$ 100 fs in pentacene \cite{wilson2011ultrafast, zimmerman2010singlet} and 100 ps in tetracene \cite{burdett2010excited, wu2014singlet, wilson2013temperature}, leading to a significant broadening. While the calculation of the singlet-fission decay rate is beyond the scope of this work, we can estimate from literature a broadening of approximately 15 meV in pentacene \cite{berkelbach2014microscopic}.
Finally, we define the non-radiative phonon-induced broadening, $\Gamma^{\mu}_\xi$. Since we are considering the low density regime, exciton-exciton scattering can be neglected \cite{perea2019exciton}. By utilising the second-order Born-Markov approximation \cite{selig2016excitonic, brem2018exciton}, we derive the exciton-phonon scattering rate for  the investigated OSCs, cf. the Supplementary Information for details.

In addition to the optical absorption, we are interested in the exciton dynamics in an oligoacene OSC. Typically this is probed via time-resolved photoluminescence (PL) spectra. 
In a similar way to the absorption spectra, we derive an Elliot-like formula for PL by using the semiconductor luminescence equation2s
\cite{kira2006many, brem2020phonon}
\begin{align}
    I^\sigma_\text{PL}(\omega)  &=  \dfrac{2}{\hbar} \sum_{\mu, \xi} \text{Im}\dfrac{N_{\mu\xi}|{\tilde{M}}^{\sigma }_{\mu\xi}|^2}{E^b_{\xi\mu} - \hbar\omega - i(\gamma_{\xi \sigma}^\mu + \Gamma^{\mu}_\xi)}
\end{align}
which differs from $I^\sigma_\text{abs}(\omega) $ by the exciton occupation $N_{\mu\xi}$. The latter can be either approximated using a thermalized Boltzmann distribution, $N_{\mu\xi} = \exp(-E^b_{\xi\mu}/ k_B T)$, or it can be calculated using the full exciton dynamics, describing the evolution of the exciton occupation $\partial_tN_{\mu\xi}$. In order to calculate the time-dependent exciton population, we derive a system of semiconductor Bloch equations \cite{selig2016excitonic,brem2018exciton}, taking into account exciton-phonon and exciton-light interactions. The resulting Boltzmann scattering equation determines exciton relaxation via phonons. We take into account both optical and acoustic phonons, arising from inter- and intramolecular phonon modes \cite{schweicher2019chasing, kamencek2022discovering}, as outlined in the Supplementary Information.

\begin{figure}[!t]
    \centering
    \includegraphics[width=0.95\linewidth]{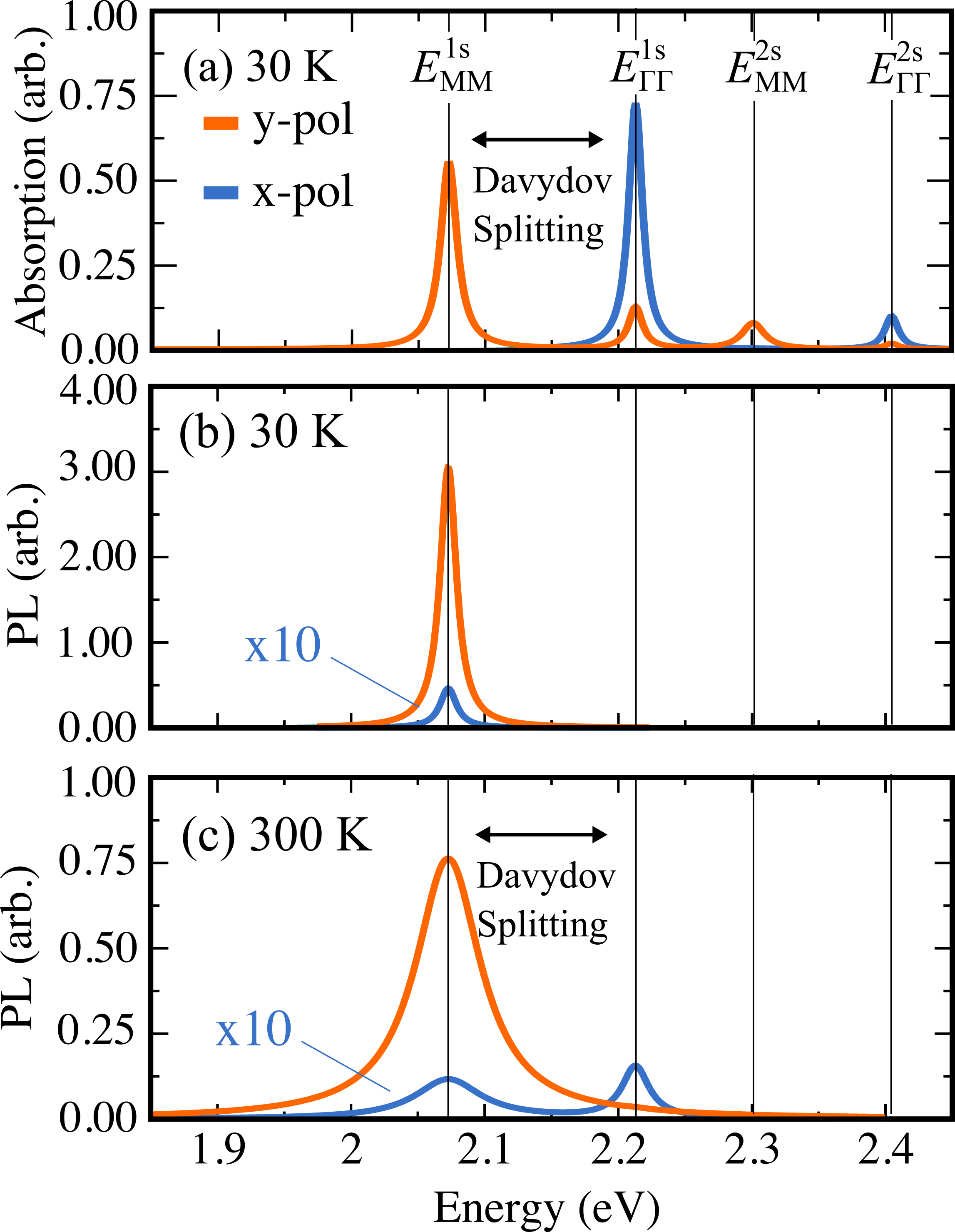}
    \caption{(a) Calculated absorption spectra of a pentacene crystal at 30 K  for y (orange) and x (blue) polarised light. Distinct peaks from the lowest 1s states from both  MM and $\Gamma\Gamma$ excitons are clearly visible and are indicated by the vertical lines. The separation between these peaks corresponds to the Davydov splitting.  The corresponding  calculated PL spectra are shown in (b) at 30 K and (c) at 300 K for y- and x-polarised light.  We scale the  x-polarised emission by a factor of 10 for clarity. The numerical aperture, taken from the experimental setup, is 0.42.}
    \label{fig:my_label2}
\end{figure}

\subsection{Exciton Optics}

Due to their unique excitonic landscape, oligoacene crystals possess a distinctive optical response. In particular, the absorption spectra of tetracene and pentacene show a clear dependence on the optical polarisation, owing to the presence of the two Davydov excitons. In Fig. \ref{fig:my_label2} (a), we show the calculated absorption spectrum  of a pentacene OSC at 30 K. The lower-energy Davydov state (the MM exciton) possesses a y-polarised optical dipole (cf. the inset in Fig 1(b)), hence an intense peak at the MM-1s resonance energy is observed for y-polarised light (orange).  There is no such peak for the orthogonal x-polarised absorption (blue) at the MM resonance. For x-polarised light, the dominant absorption  peak stems from the $\Gamma\Gamma$-1s exciton. Since the MM and $\Gamma\Gamma$ optical dipoles are not orthogonal (cf. the inset in Fig 1(b)),  a small signature at $\Gamma\Gamma$ is still observed for y-polarised light. This polarisation dependence and splitting is characteristic of the Davydov splitting in these crystals \cite{cocchi2018polarized}, which we find to be approximately 140 meV in magnitude. In addition to  1s-excitons, we also find signatures from higher 2s exciton states. The excitonic wavefunction of these states is less localised leading to a less efficient optical matrix element and hence less intense resonances. 

In Figs. 2 (b)-(c), we show the corresponding PL spectra of the pentacene crystal at 30 K and 300 K, respectively.  For now, we assume that our system is thermalised into a Boltzmann distribution, resulting in a much lower occupation of the upper Davydov and 2s excitons. At the low temperature of 30 K this means that  we only observe an emission peak stemming from the energetically lowest MM-1s exciton for both y- and x-polarisations. Here, it is important to clarify that we assume a non-zero numerical aperture, in order to better represent a typical experimental situation. The non-zero out of plane optical dipole leads to a non-negligible coupling between the MM-1s exciton and the photons emitted non-orthogonally to the plane for all in-plane linear polarisations. As a result, the MM exciton couples with the x-polarised light unlike in the absorption case, where we did not consider non-orthogonal incidence (Fig. 2 (a)). The contribution from the x-polarised emission is, however, rather small, owing to the weak optical matrix element and little occupation of the $\Gamma \Gamma$ exciton (note the factor of 10 in Fig. 2 (a)).

At 300 K, the upper $\Gamma\Gamma$ exciton becomes more significantly occupied and as a result we observe a PL signature at the $\Gamma \Gamma$ exciton for x-polarised light (blue line in Fig. 2 (c)). The relative intensity of the peaks is dictated by the relative occupation of the two excitons, the optical matrix element and the broadening.  For $y$-polarised light, the intensity of the MM exciton is significantly larger and, combined with the larger occupation of the MM exciton, surpasses by far the contribution of the $\Gamma \Gamma$ exciton (orange line). Furthermore, at higher temperatures, the phonon population is larger, leading to an increased exciton-phonon interaction and hence much broader exciton linewidths, as shown when comparing Figs. 2(b) and (c). Here, we do not take into account either the Stokes shift nor the temperature-dependent red shift in the exciton resonance \cite{koch2006evidence}, which has been measured previously to be about 20 meV in pentacene thin films \cite{helzel2011temperature} for both Davydov components. 
Our theoretical model does not only show a good agreement with previous theoretical and experimental studies of optical absorption \cite{cocchi2018polarized, schuster2007exciton}, but it also reveals an interplay between the  polarisation and temperature dependence of PL spectra \cite{camposeo2010polarized}. In particular, the higher optical matrix element of the $\Gamma\Gamma$ exciton at x-polarisation competes with the lower thermal occupation, which determines the dominant spectral signatures.

\subsection{Exciton Dynamics}
Exciton dynamics plays a crucial role in the optical response of OSC crystals. However, the dynamics particularly of singlet excitons is poorly understood, with exciton-phonon scattering and dark-exciton states both likely to play a critical role. Our microscopic many-particle theory is particularly suited to describe the ultrafast relaxation dynamics of optically excited excitons \cite{selig2016excitonic,brem2018exciton}.
By solving the Boltzmann scattering equation for excitons (cf. Supplementary Information), we obtain microscopic access to the time evolution of the population within the light cone of the bright MM and $\Gamma\Gamma$ excitons as well as the momentum-dark M$\Gamma$ and $\Gamma$M excitons, cf. Fig. 3 (a). Since the excitation is x-polarised, coherent $\Gamma\Gamma$ excitons are initially created by the incident optical pulse (not shown), which then quickly converts into an incoherent exciton population via polarisation-population transfer \cite{brem2018exciton, selig2018dark}.  This leads to a sharp increase  in the $\Gamma \Gamma$ population after approx. 100 fs, appearing as an initial large population at time $t=0$ fs, shown by the yellow curve in Fig 3(a).  Following this, intraband scattering, mediated by acoustic phonons, spreads out the momentum distribution within the band on an ultrafast timescale (sub ps),  lowering the population of the $\Gamma \Gamma$ exciton within the light cone. In addition, interband scattering mediated by optical phonons allows for relaxation between bands, with scattering pathways shown in the inset of Fig 3(a). 

Interestingly, the small energy separation between the $\Gamma\Gamma$ and $\Gamma$M excitons permits rapid exciton relaxation from the $\Gamma\Gamma$  to $\Gamma$M state via optical phonon emission. Therefore the $\Gamma$M population (blue) begins to increase within a very  short timescale (ps) following a rapid depopulation of the $\Gamma \Gamma$ exciton. In contrast, the large energy separation and relative flatness of the exciton bands leads to a phonon bottleneck effect, slowing down the population transfer of excitons from the $\Gamma\Gamma$ and $\Gamma$M  to the lower lying M$\Gamma$ and MM states (ns), respectively (cf. the inset in Fig 3 (a)). As a result we find a much slower transfer of exciton population from the higher energy states to the lower ones, resulting in a relaxation occurring on the timescale of ns. This is demonstrated by the slow increase of the MM (red) and M$\Gamma$  (light blue) populations in Fig 3(a).
While the timescale of this population transfer is dependent on the specific material parameters that might vary for different molecular crystals, our results are fairly general, demonstrating that dark excitons are crucial scattering pathways in the exciton relaxation from the upper to the lower Davydov state in herringbone oligoacene semiconductor crystals. 

\begin{figure}[!t]
    \centering
    \includegraphics[width=0.95\linewidth]{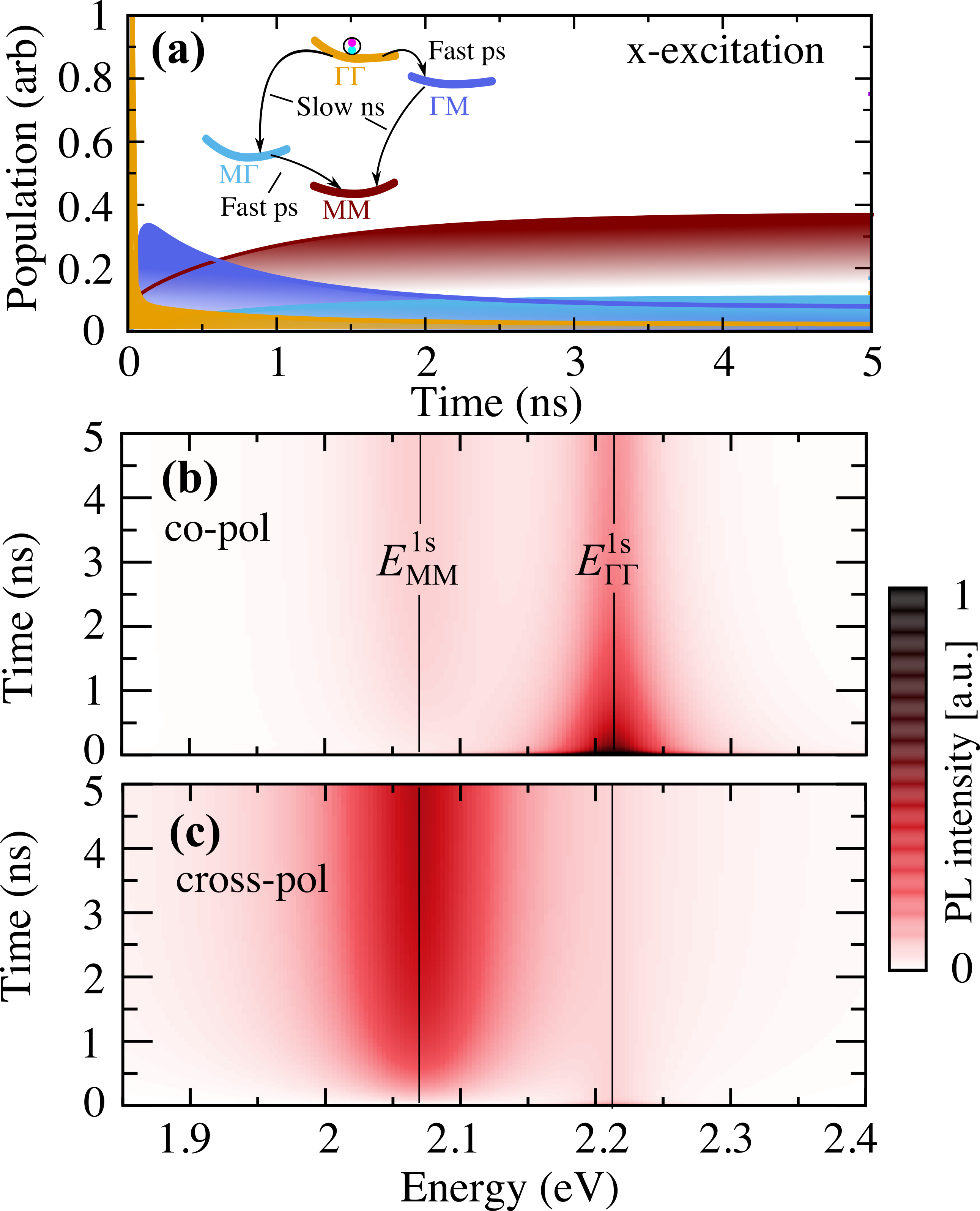}
    \caption{
    (a) Full timescale exciton dynamics of the incoherent exciton population within the light-cone of pentacene crystals. The initial excitation is an  x-polarised 100 fs wide Gaussian pulse.  The inset shows a schematic of the exciton relaxation pathway. The final state is the same for y-polarised excitation (not shown here), as the final steady state is independent of the excitation polarisation. Time-resolved PL of a pentacene crystal at 300 K for (b) co- and (c) cross-polarised light.  }
    \label{fig:my_label4}
\end{figure}

 In Fig 3 (b) and (c), we show time-resolved PL colour maps at 300 K. We assume an x-polarised excitation (driving the $\Gamma\Gamma$ exciton) and calculate the PL intensity for co- ($x$) and cross-polarised ($y$) emission, respectively.
 In the co-polarised case (Fig 3 (b)) the emitted photons are also x-polarised, hence the PL signal is dominated by the $\Gamma\Gamma$ excitons at early times. As time passes, MM excitons become the most populated states, leading to a clear emission peak at the MM resonance. Here, it is natural to compare the intensity at 5 ns to the data presented in Fig 2 (c). We notice immediately that the $\Gamma \Gamma$ exciton peak is larger relative to the MM peak in the dynamics calculation, suggesting the thermalized Boltzmann assumption used for Fig. 2(c), underestimates the exciton population at larger energies. This discrepancy is due to the phonon bottleneck in the system preventing efficient relaxation from the upper to the lower Davydov state. This bottleneck arises since phonons can not provide sufficient energy transfer between the higher energy $\Gamma\Gamma$/$\Gamma$M and the lower M$\Gamma$/MM exctions, stifling the relaxation process.  As the bottleneck effect is strongly dependent on the actual Davydov splitting, systems with smaller splittings are less likely to display phonon-bottleneck effects, as it will turn out to be the case in tetracene crystals.
 Finally, in the cross-polarised case (Fig. 3 (c)),  corresponding to $y$-polarised emission, the optical matrix element of the MM exciton is the largest. Hence, we observe the emergence of a dominant MM exciton peak after about 100 ps, corresponding to the population transfer from the $\Gamma\Gamma$ to the MM exciton.  
  
 Note that in pentacene crystals, rapid conversion of singlet excitons into triplets takes place ($\sim$100 fs) \cite{congreve2013external, thorsmolle2009morphology}. This practically limits the PL lifetime and hinders measurements on the dynamics of these systems. Despite this, signatures of the upper Davydov peak have been previously observed in pentacene \cite{he2005fundamental, tayebjee2013exciton, rinn2017interfacial} which is characterized by a large Davydov splitting, indicating a non-Boltzmann distribution of excitons.

\subsection{Theory-experiment comparison}
Compared to pentacene,  singlet-fission occurs on a significantly lower timescale in tetracene and requires thermal activation \cite{thorsmolle2009morphology} and is believed to be phonon-driven \cite{morrison2017evidence}. The singlet-exciton relaxation is therefore likely to be more important for device operation in tetracene crystals. Furthermore, tetracene is more suited for time-resolved PL studies, as exciton relaxation is easier to observe experimentally than in pentacene, where the emission is rapidly quenched by singlet fission \cite{he2005fundamental, tayebjee2013exciton}. Furthermore, the critical role of phonons in the singlet-fission process \cite{morrison2017evidence} in tetracene suggests that the energetic landscape of the singlet excitons and their relative occupation is important for understanding these processes.
Therefore, now we present experimental absorption and PL measurements on a high-quality tetracene crystal and directly compare the data to our theoretical predictions.  

\begin{figure}[!t]
    \centering
    \includegraphics[width=0.99\linewidth]{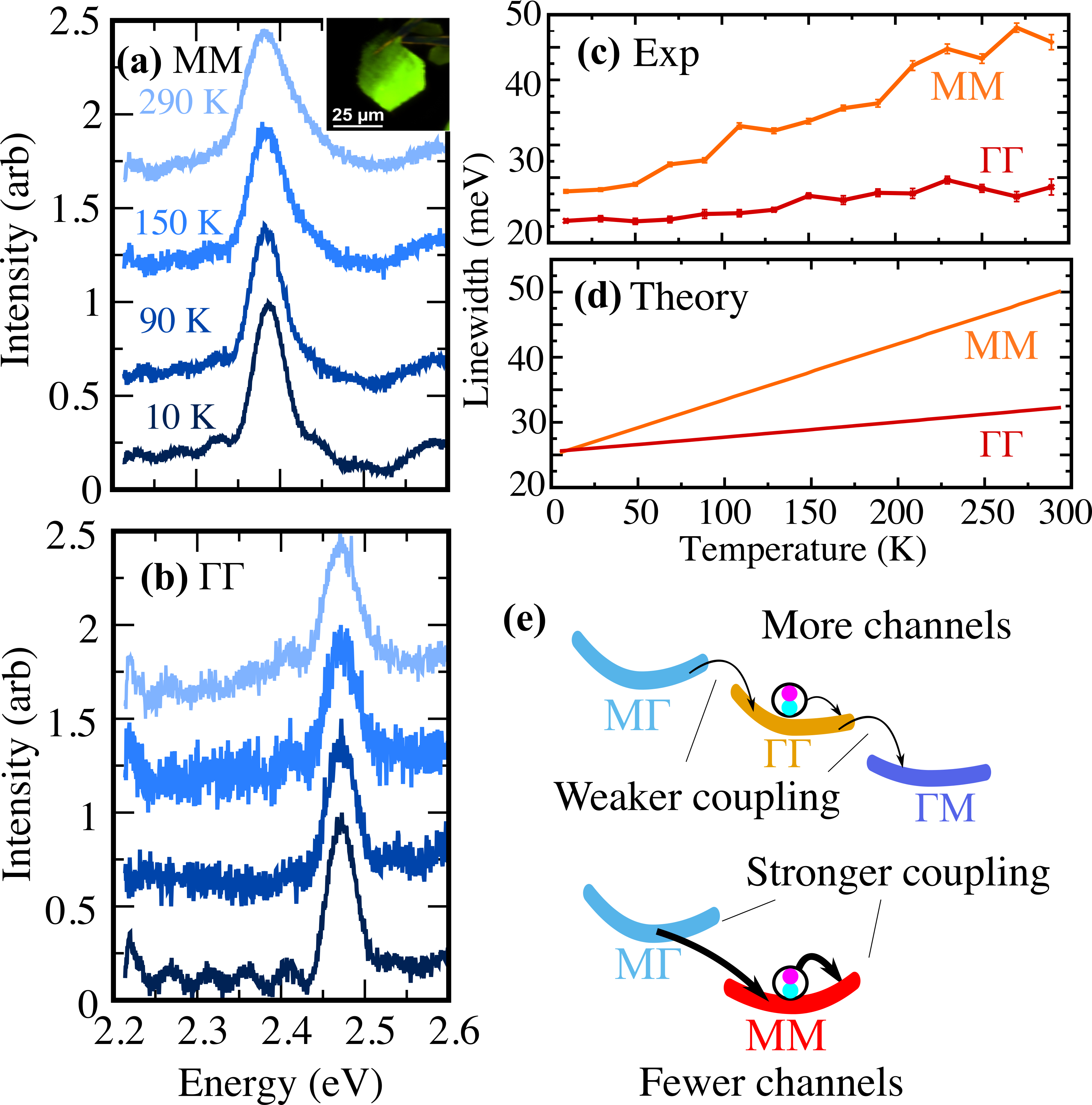}
    \caption{Experimental absorption spectra on a tetracene crystal at different temperatures for (a) y-polarised and (b) x-polarised light probing the MM and $\Gamma\Gamma$ excitons, respectively. Inset in (a) shows an optical micrograph of the tetracene sample. (c) Experimental temperature-dependence of the absorption linewidth (with error bars) for both the MM and $\Gamma\Gamma$ excitons in a tetracene crystal. (d) Theoretical fit by tuning the electron-phonon coupling strength of tetracene crystals. (e) Schematic showing the key phonon-scattering channels determining the exciton linewidth, suggesting that the MM exciton should posses stronger exciton-phonon coupling.}
    \label{fig:my_label5}
\end{figure}

Tetracene is structurally and therefore optically very similar to pentacene. It also forms a herringbone structure, giving rise to distinct Davydov components in the optical response. In tetracene, like in pentacene, the electronic dispersion possesses two clear direct band gaps, at MM and $\Gamma\Gamma$. The main differences between the two structures are of quantitative nature, resulting from the slightly different band gap, intermolecular coupling, dipole orientation and dielectric screening.
We show absorption measurements on tetracene  for different temperature in Fig. 4, probing the (a) MM and (b) $\Gamma\Gamma$  excitons through y-polarised and x-polarised light, respectively. Similarly to pentacene, we observe, a clear Davydov splitting, however this time only $\sim$ 70 meV. We find that the broadening of the MM exciton increases much more rapidly with temperature than for the $\Gamma \Gamma$ exciton. The general temperature dependence of the linewidth is due to the increased exciton-phonon scattering at higher temperatures \cite{brem2019intrinsic}.
 Therefore, we can exploit these absorption measurements to obtain important information about the exciton-phonon scattering rate. In particular, in Fig. 4, we extract the absorption linewidth from (a) and (b), and compare the experimental linewdith (c)  with theoretical predictions (d) for the MM and $\Gamma\Gamma$ excitons at different temperatures.   Clearly, we see that the linewidth of the MM exciton is significantly larger and increases more rapidly with temperature than for the $\Gamma\Gamma$ exciton. This is somewhat surprising as the scattering rate is typically larger for higher energy excitons, due to the higher number of scattering channels  (cf. Fig 4 (e)). We attribute this difference to a more efficient exciton-phonon coupling for the MM exciton than the $\Gamma \Gamma$ exciton \cite{brem2019intrinsic}. This broadening is driven by both intraband and interband transitions, taking into account both optical and acoustic phonons (see Supplementary Information).

In Fig 5 (a), we show the experimental PL spectra at 30 K for y-polarised (blue) and x-polarised (orange) light. As in the case of pentacene at low temperature (cf. Fig 2 (b)) the signal is dominated by the MM exciton for both polarisations as the upper $\Gamma\Gamma$ exciton is far less occupied. The polarisation behaviour and relative intensities are well reproduced in our calculations, as shown in Fig 5 (b). In the experimental data, we see a pronounced asymmetry in the PL spectra, which we attribute to the vibronic coupling in molecular crystals \cite{burdett2010excited, franck1926elementary, condon1926theory} that has not been taken into account in the simulations. 
\begin{figure}[!t]
    \centering
    \includegraphics[width=0.99\linewidth]{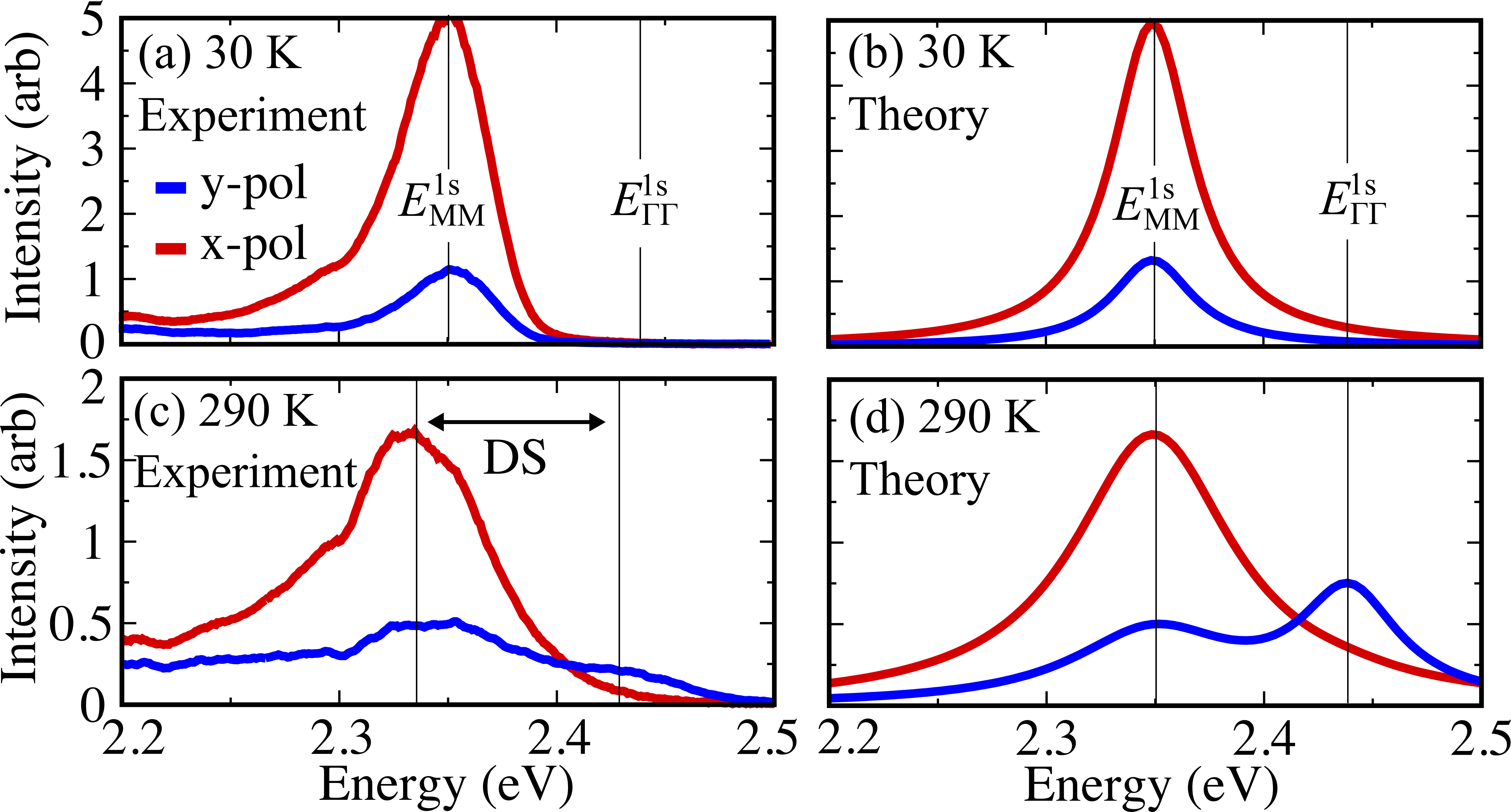}
    \caption{Direct comparison of experimental [(a), (c)] and theoretical PL spectra [(b), (d)] of a tetracene crystal for the y- and x-polarised excitation at 30 and 290 K, respectively. The Davydov splitting (DS) is indicated with an arrow.   }
    \label{fig:my_label5}
\end{figure}

Compared to the pentacene case, the Davydov splitting is now only 70 meV and hence the exciton population of the upper Davydov state is significant at 300 K. Therefore, in Figure 5(c) we observe a clear experimental PL peak for y-polarised emission originating from the $\Gamma \Gamma$ exciton, manifesting as a pronounced shoulder at around 2.43 eV, in addition to the signature from the MM exciton. However for x-polarised emission, the signal from the $\Gamma \Gamma$ exciton vanishes, owing to the reduced population and lower optical matrix element.  The same features are recovered in our theoretical calculation (Fig. 5(d)), again outlining the strength of our model. For example, at 30 K, we find that the peak ratio between the y-polarised and x-polarised intensity stemming from the MM exciton is approximately 0.25, in both experiment and theory. At 300 K, we find clearly the emergence of the upper Davydov peak in both theory and experiment. 

The theoretical PL spectra are calculated using the exciton population 100 ps after the excitation pulse, where the system has reached a thermalised exciton distribution. This allows us to test whether phonon-bottleneck effects are observed. The results using the Boltzmann distribution for the exciton population, like those calculated in Figs. 2(b,c), give quantitatively nearly identical results, suggesting that there is no measurable phonon-bottleneck effect in tetracene molecular crystals (in contrast to the pentacene case). We attribute this to the smaller energetic separation between the Davydov components. This smaller energy separation allows energy conservation of the interband exciton-phonon scattering process (cf. Fig 1 (c)) to be more easily fulfilled increasing the relaxation efficiency. 
While we find excellent overall agreement, the intensity ratios between the MM and $\Gamma \Gamma$ excitons  differ quantitatively between theory and experiment at 300 K, likely due to photon reabsorption effects \cite{burdett2012quantum} and extrinsic broadening mechanisms, such as defects \cite{mcanally2017defects}.

\section{Conclusion}
In conclusion, we have performed a joint theoretical and experimental study to explore  optical and dynamical properties of oligoacene herringbone crystals. We combine the fabrication of high-quality crystals and polarization- and temperature-resolved PL measurements with a highly-predictive microscopic many-particle theory. We successfully recover the Davydov splitting in pentacene, and show polarisation and temperature dependence of the absorption and PL spectra, with distinct signatures from the two Davydov peaks at different polarisations and temperatures. Using a microscopic model for the exciton phonon-scattering, we calculate the exciton dynamics and demonstrate the emergence of a phonon-bottleneck in pentacene crystals, arising from the flatness and energy separation of the bright and dark exciton bands. This shows that phonons not only cause a broadening of absorption peaks but also affect the population dynamics of Davydov components. We demonstrate that dark excitons represent a crucial scattering-channel for exciton relaxation in these systems. We directly compare our microscopic and material-specific model to experimental absorption and PL measurements on high-quality tetracene. We reproduce the same PL peak-intensity ratios between the two Davydov peaks at different temperatures and polarisations as well as excitonic linewidths for both  absorption and PL spectra. Our results represent an important advance in microscopic understanding of optics and dynamics properties of organic semiconductor crystals paving the way for future studies on this technologically promising material platform.

\noindent \textbf{Supplementary Information}
Supplementary Information: Figure outlining the determination of the crystallographic structure and PL/absorption comparison for different film thickness. Details on the fabrication, spectrocscopy and theoretical model.

\noindent \textbf{Acknowledgements}
For the spectroscopic studies we thank Prof. Martin Koch for providing access to experimental equipment. We thank Prof. Michael Rohlfing for fruitful discussions.  This project has received funding from Deutsche Forschungsgemeinschaft via CRC 1083 (project A02, B09 and B10) and the European Unions Horizon 2020 research and innovation programme under grant agreement no. 881603 (Graphene Flagship).

\bibliographystyle{achemso}
\providecommand{\latin}[1]{#1}
\makeatletter
\providecommand{\doi}
  {\begingroup\let\do\@makeother\dospecials
  \catcode`\{=1 \catcode`\}=2 \doi@aux}
\providecommand{\doi@aux}[1]{\endgroup\texttt{#1}}
\makeatother
\providecommand*\mcitethebibliography{\thebibliography}
\csname @ifundefined\endcsname{endmcitethebibliography}
  {\let\endmcitethebibliography\endthebibliography}{}

\end{document}